\documentclass[aps,prl,twocolumn,superscriptaddress,nobibnotes]{revtex4-1}

\usepackage{graphicx}
\usepackage{dcolumn}
\usepackage{bm}

\usepackage[utf8]{inputenc}
\usepackage[T1]{fontenc}
\usepackage{mathptmx}
\usepackage{etoolbox}
\usepackage{float}
\usepackage{xcolor}
\begin{document}


\title{Free-Space Quantum Key Distribution with Single Photons from Defects in Hexagonal Boron Nitride} 



\author{Çağlar Samaner}
\thanks{These authors contributed equally}
\affiliation{Department of Physics, İzmir Institute of Technology, İzmir, 35430, Turkey}

\author{Serkan Paçal}
\thanks{These authors contributed equally}
\affiliation{Department of Physics, İzmir Institute of Technology, İzmir, 35430, Turkey}

\author{Görkem Mutlu}
\affiliation{Department of Physics, İzmir Institute of Technology, İzmir, 35430, Turkey}

\author{Kıvanç Uyanık}
\affiliation{Department of Physics, Gazi University, Ankara, 06500, Turkey}

\author{Serkan Ateş}
\email[]{Corresponding author: serkanates@iyte.edu.tr}
\affiliation{Department of Physics, İzmir Institute of Technology, İzmir, 35430, Turkey}


\date{\today}
\begin{abstract}
We present a proof-of-concept demonstration of free space quantum key distribution (QKD) with single photons generated from an isolated defect in hexagonal boron nitride (hBN). The source, operating at room temperature with a 10\% brightness, is integrated into a B92 protocol and a secure key rate (SKR) of 238 bps and a quantum bit error rate (QBER) of 8.95\% are achieved with 1 MHz clock rate. The effect of temporal filtering of detected photons on the performance of QKD parameters is also studied. We believe that our results will accelerate the work on improving the performance of optically active defects in hBN and their use in high-performance practical QKD systems. 

\end{abstract}

\pacs{}

\maketitle 

Quantum key distribution (QKD) is one of most developed applications within quantum information technologies. The first proposed QKD protocol, known as BB84 \citep{Bennett1984}, relies on polarization of photons for encoding information. Among several approaches, using weak coherent pulses as the light source provides a practical approach to realize a QKD system, although it is inefficient~\citep{Bennett1992a} and has possible security weaknesses under photon number splitting attacks (PNS)~\citep{Brassard2000}. These issues are addressed by using decoy states protocols~\citep{Hwang2003,Lo2005}, and integrated photonic and electronic circuits are already realized with decoy state QKD systems\citep{Wang2020, Paraiso2021}. However, ultimate security and long distance operation still requires true single photon sources (SPSs) with high efficiency. To date, several SPSs have been used in different QKD demonstrations, such as optically~\citep{Waks2002,Intallura2007,Collins2010,Takemoto2010}, or electrically~\citep{Heindel2012,Rau2014} excited semiconductor quantum dots (QDs) and color centers in diamond~\citep{Beveratos2002,Alleaume2004,Leifgen2014}. Although semiconductor QDs have the advantage of easier integration with cavity based systems and electrical operation, they require cryogenic temperatures to operate efficiently. Alternatively, color centers of diamond nanocrystals are operable in room temperature with bright emission. Further, they can also be implemented in nano-photonic structures for applications in integrated photonic technologies \citep{Bhaskar2020}. In addition to these SPSs, a recent study shows that single dibenzoterrylene molecules embedded in nano-crystals are also good candidates for QKD systems due to their high efficiency at room temperature~\citep{Murtaza2022}. On the other hand, there is a growing interest on entanglement based QKD systems~\citep{Ekert1991,Bennett1992_ent} that mostly use nonlinear crystals~\citep{Jennewein2000} or semiconductor quantum dots~\citep{Dzurnak2015,Schimpf2021,Basset2021} for efficient generation of entangled photon pairs. 

Despite all the efforts on the above-mentioned SPSs, the interest has been grown in quantum emitters in hBN,  known as a two dimensional (2D) indirect bandgap semiconductor~\citep{Cassabois2016}. Thanks to its wide bandgap ($\approx 6~eV$), hBN hosts several types of optically active defects~\citep{Weston2018} that generate efficient single photon emission from cryogenic temperatures~\citep{Tran2015, Martinez2016,Grosso2020,Jungwirth2016} up to 800 K~\citep{Kianinia2017} over a wide spectral range from UV~\citep{Bourrellier2016}  to NIR~\citep{Tran2017}. The fact that hBN hosts bright and optically stable emitters, in addition to its scalability, owing to its 2D nature, its a great candidate for the applications towards the quantum technologies at room temperature, compared to the alternatives. \citep{Shaik2021,White2021,Kianinia2022,Kubanek2022}. 

Here, we present, to the best of our knowledge, the first proof-of-concept demonstration of B92-based QKD~\citep{Bennett1992} with single photons generated from an isolated defect in hBN. Brightness and purity of zero phonon-line (ZPL) emission of the defect are characterized under pulse excitation conditions at room temperature after which it is implemented in a QKD system where encoding is realized at 1~MHz clock rate. Secure key rate (SKR) and quantum bit error rate (QBER) are also optimized via temporal filtering of detected photons. We anticipate that our results will stimulate the implementation of these emitters for various applications in quantum technologies.

\begin{figure*}[!htp]
\includegraphics[width=0.98\linewidth]{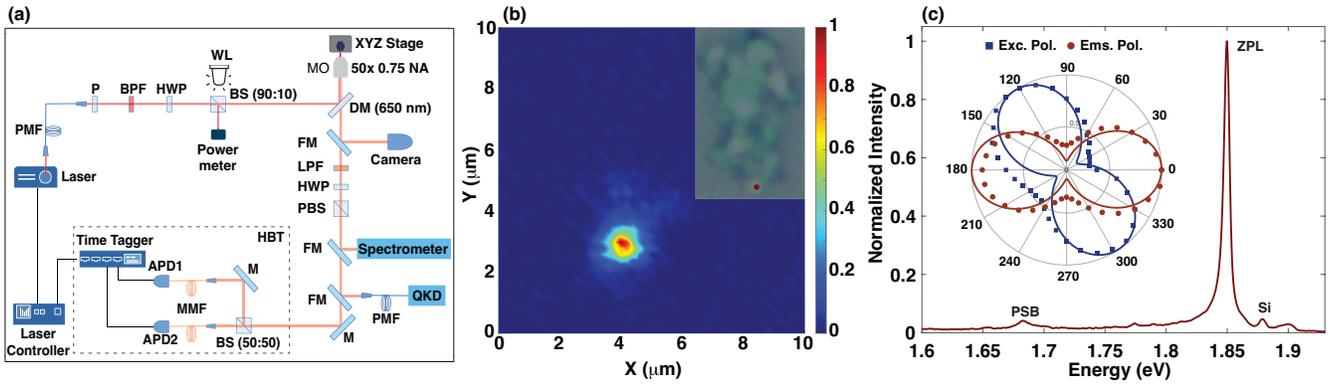}
\caption{\label{fig:fig1} (a) Schematic representation of confocal micro-PL setup that is used to characterize the optical properties of single defects in hBN. Collected emission is spectrally analyzed by a spectrometer, and its single photon nature is measured with an HBT interferometer. Spectrally filtered and polarized single photon emission is guided to the QKD system via a polarization maintaining single mode fiber. (b) PL map of a bulk hBN with a bright localized emission. Inset shows the optical image of the studied hBN structure. (c) PL spectrum of the isolated defect taken under 1 MHz repetition rate. The inset shows the excitation (square) and emission (circle) polarization dependent intensity of ZPL emission.}
\end{figure*}

Schematic of the experimental setup that is used to characterize the optical properties of defects in hBN is given in Fig.~\ref{fig:fig1}(a). A 637 nm pulsed laser is used for the excitation that passes through a polarizer and a half-wave plate (HWP) for polarization manipulation. A high numerical aperture microscope objective (NA = 0.75) is used to focus the laser light and to collect the emission from the sample, which is placed on a high-resolution XYZ stage for precise positioning and for the scanning purposes. A 650 nm long pass dichroic mirror is used to direct the laser light onto the sample and transmit the emission to the detection port efficiently. An additional 650 nm long pass filter is placed on the detection port for a stronger filtering of laser light. Collected emission is then passed through a HWP and a polarizing beamsplitter (PBS) for polarization analysis and directed either to a spectrometer for spectral analysis or to a Hanbury-Brown and Twiss (HBT) interferometer for photon correlation measurements. The interferometer has a 50/50 beamsplitter and two ID120 single-photon detectors. Finally, pre-characterized emission is guided to the QKD setup by a polarization-maintaining single mode fiber, which also acts as a pinhole to spatially filter the ZPL emission. Details of QKD setup is described below.

Multilayer hBN structures are obtained in a solution from Graphene Supermarket and drop-casted on SiO2/Si substrate. All data presented in this work belongs to a single defect in hBN operating at room temperature.  Photoluminescence map of the investigated structure is given in Fig.~\ref{fig:fig1}(b). An optical image of this hBN flake is shown as an inset. The localized bright emission observed in the map is due to an isolated defect with the emission spectrum (recorded under 1 MHz laser repetition rate) shown in Fig.~\ref{fig:fig1}(c). The defect has a sharp zero-phonon line (ZPL) emission at 1.848 eV (671 nm) with a characteristic optical phonon sideband (PSB) at 1.683 eV (736 nm).  Energy difference between the ZPL and the PSB is about 165 meV which matches perfectly with the energy of the high-energy Raman active phonon mode of the host hBN. As observed, the ZPL emission from the defect dominates strongly in the whole spectrum, indicating a large Debye-Waller factor~\citep{Ari2018}. The small peak at 1.878 eV is due to the Raman scattering from the silicon substrate. The inset of Fig.~\ref{fig:fig1}(c) shows the absorption and emission polarization of the ZPL with a high degree of visibility as expected for a single dipole~\cite{Jungwirth2017}. The strong linear polarization of the ZPL emission observed here is desirable for QKD implementations that use the source with the highest efficiency. 

\begin{figure}[!hbp]
\includegraphics[width=0.97\linewidth]{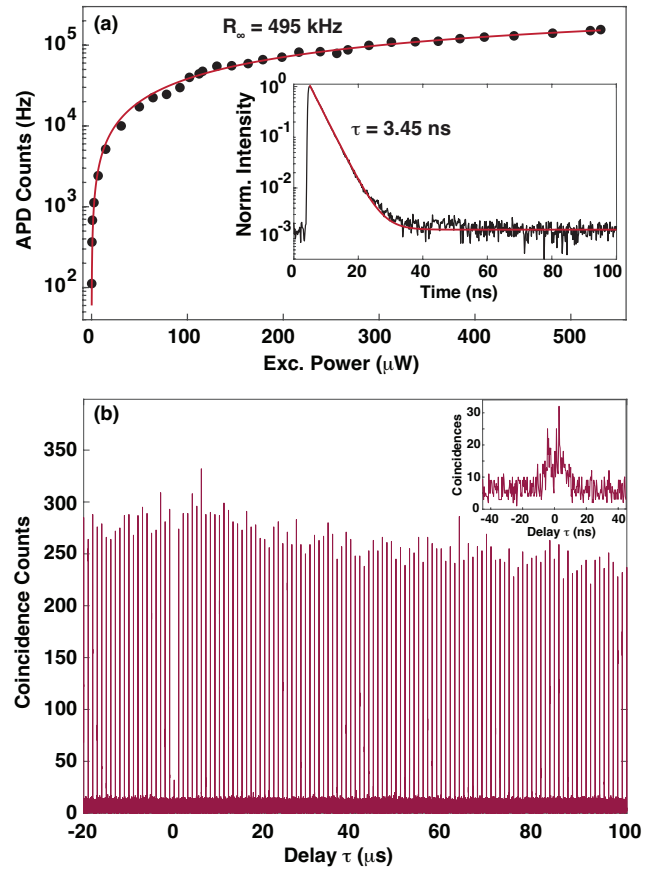}
\caption{\label{fig:fig2} (a) Excitation power dependence of ZPL emission (spectrally filtered) measured with an APD under 40 MHz repetition rate. (inset) Time-resolved PL of the defect with a decay time of 3.45 ns. (b) Second-order photon correlation function  for the same emitter taken under 1 MHz repetition rate. The anti-bunching value is estimated as $g^{(2)}(0) = 0.12 \pm 0.03$. The inset shows the magnified zero time delay region for clarity. 
}
\end{figure}

Brightness is another important property of the emitters for QKD applications. To quantify the brightness of the same emitter, excitation power dependent saturation measurement is performed on the spectrally filtered (FB670-10, Thorlabs) ZPL emission with a fiber-coupled single photon detector. Figure~\ref{fig:fig2}(a) shows the result of this measurement taken under 40 MHz repetition rate of the laser and the solid line represents the fit using  $R = (R_{\infty}*P)/(P+P_{sat})$ function obtained from three-level model of defects in hBN~\citep{Tran2015}. Here, $R_{\infty}$ and $P_{sat}$ are the maximum emission rate and excitation power at which the intensity saturates. Although the highest measured count rate is 150~kHz under maximum available excitation power, fit result indicates a maximum achievable emission rate of 495~kHz under stronger excitation conditions. Considering all the imperfections of the components used for this measurement (detector efficiency, coupling into multimode fiber, efficiency of bandpass filter, the transmission of dichroic mirror and the efficiency of several optics over the path), a count rate of 1.2~MHz is achieved after the microscope objective. This corresponds to a 3\% of collection efficiency under the highest excitation power of the laser with 40~MHz repetition rate, which can be as high as 10\% if saturation could be achieved. Here, we have assumed a 100\% quantum yield of the emitter. However, direct measurement of quantum yield for similar types of defects in hBN were reported to be about 40\% \citep{Nikolay2019}. Collection efficiency of the ZPL emission can be further improved by either using solid-immersion lenses~\citep{Zeng2022} or coupling the emitters to nanophotonic \citep{Froech2020} or plasmonic structures \citep{Nguyen2018}.

In addition to the saturation, the dynamics of the ZPL emission is measured using time-correlated single photon counting method. Inset of Fig.~\ref{fig:fig2}(a) shows the result of such an experiment under 10 MHz excitation rate. A single exponential decay with a typical value of $3.45 \pm 0.05~ns$ is observed for the studied defect~\citep{Tran2015}. We also examined the purity of the emitted photons by measuring the second-order photon correlation function, $g^{2}(\tau)$, using the HBT interferometer. The result of such an experiment performed on the spectrally filtered ZPL emission is shown in Fig.~\ref{fig:fig2}(b). While the number of coincidences decrease exponentially over long time delays as expected for a three-level emitter, a clear anti-bunching is observed at zero time delay with a value of $g^{2}(0)=0.12$. Inset shows the magnified data around zero time delay where two sharp peaks are considered as cross-talk between APDs. This experiment is performed under 1 MHz excitation rate, which is also used for the demonstration of QKD with the same emitter.  To this end, the linearly polarized ZPL emission is coupled to a polarization-maintaining fiber and directed to the QKD setup.
\begin{figure}[!tp]
\includegraphics[width=0.95\linewidth]{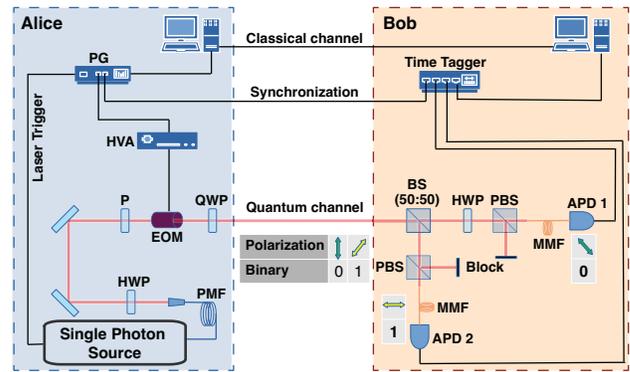}
\caption{\label{fig:fig3} Experimental setup of B92-based free-space QKD system. Alice Side: One channel of the pulse generator drives the single photon source system at 1 MHz repetition rate and simultaneously sends the reconciliation signal to the time tagging module, which is used for synchronization of the system. The other channel sends a periodic pattern signal to the EOM. The modulator controls the polarization of the photon to be V (0) or +45 (1) according to the incoming signal. Bob Side: Beam splitter (BS) randomly directs the incoming photons to either APD1 or APD2 for polarization analysis. APD1 detects only if the photon has -45 (0) polarization while APD2 detects if it has H (1) polarization and the time of each detection events are recorded by time tagging module for reconciliation process.
}
\end{figure}

The schematic of the free-space QKD system is shown in Fig.~\ref{fig:fig3}. Here, a B92-based protocol is implemented due to its simplicity and the availability of the equipment with a cost of half key rate efficiency compared to the well known BB84 protocol. The transmitter (Alice) uses a pulse generator to trigger the excitation laser (inside the single photon source box) and to feed the high voltage amplifier (New Focus HVA 3211) that drives the electro-optic amplitude modulator (EOM, New Focus 4102) utilized for polarization encoding of single photons. Due to the capacity of the high voltage amplifier, the maximum repetition rate of the entire system is limited to 1 MHz. Single photon emission is passed through a half-wave plate and a polarizer to set vertical polarization input with a high extinction ratio before entering the EOM. In our implementation of the protocol, vertical and +45 degree polarizations are encoded on single photons, which are obtained by driving the EOM with  0~V and 95~V, respectively, and using a quarter-wave plate (QWP) behind the EOM to convert the circular polarization output of EOM to linear polarization. In this experiment, a periodic encoding of polarizations (010101...) is used for a proof-of-concept demonstration~\citep{Rau2014}. The receiver (Bob) employs a non-polarizing 50:50 BS to choose a random measurement basis, and two PBS and a HWP to polarization selection. In addition, two fiber-coupled APDs are attached to the output ports of each PBS for detection of single photons. Finally, a time tagging module is used both to record the time of measurement events from each APD and to synchronize Alice and Bob via the common trigger signal generated by the pulse generator on the Alice side.

The demonstration of QKD is performed under 1~MHz repetition rate and at the highest achievable power (not enough to saturate the defect as discussed before) of the excitation source. A total count rate of 30~kHz after the microscope objective and about 7.8~kHz at the input of Alice is obtained under this excitation condition. The count rate is further degraded to a 400 Hz on each APD (on top of $\approx$1.5~kHz dark counts) because of all the imperfections of the QKD system (i.e., optics, EOM, detection efficiency of APDs, fiber couplings, and the 25\% efficiency of the protocol), which therefore indicates a total efficiency of the QKD system after the first lens to be about 2.67\%.

\begin{figure}[!tp]
\includegraphics[width=0.95\linewidth]{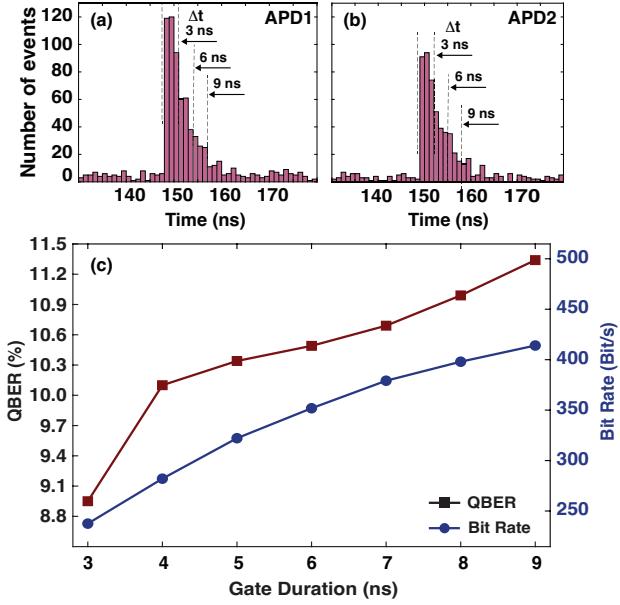}
\caption{\label{fig:fig4} (a) and (b) Histogram of relative time differences between laser trigger and detection on APDs. Each bin shows the total counts in 1 ns time interval for a total of 2.5 second of QKD demonstration. (c) Filtering duration QBER and key rates both of which increases as the selected region of detection duration increases. 
}
\end{figure}

 Figure \ref{fig:fig4} (a) and (b) show the histograms of relative time difference between a laser trigger and APD detection events that are acquired for a 2.5~s run of QKD experiment. Since the lifetime of the emitter is 3.45~ns, main contribution to the observed 148 ns delay in the histograms comes from the electronics of the system. The distribution of detection events around 148 ns delay gives a bound for SKR and QBER analysis, which is used to eliminate the effect of random dark counts of APDs outside of this bound. However, there is a trade-off between QBER and key rate as a function of temporal filtering, $\Delta t$, starting at 148 ns delay as shown in the Fig.~\ref{fig:fig4}(c). As observed, $\Delta t$ increases QBER due to the increasing contribution of dark counts. Considering the events in the 3~ns and 9~ns filtering durations, QBER and SKR are measured to be $\text{QBER}_{\text{3ns}}~=~8.95\%$, $\text{SKR}_{\text{3ns}}$~=~238~Bit/s and $\text{QBER}_{\text{9ns}}~=~11.34\%$, $\text{SKR}_{\text{9ns}}$~=~414~Bit/s, respectively. Below this limit, SKR becomes significantly low, making it inefficient for any practical purposes. Extending the temporal filtering duration increases the SKR but also brings the QBER closer to security limits. Performance optimization via similar temporal filtering was also reported for QD-based QKD systems recently~\citep{Kupko2020}. It is important to highlight that the measured key rate is limited by the maximum available power of the excitation laser. As we discussed earlier, the SKR can be increased by at least a factor of 3 if the defect is excited at its saturation. 
 
\begin{figure}[!tp]
\includegraphics[width=0.9\linewidth]{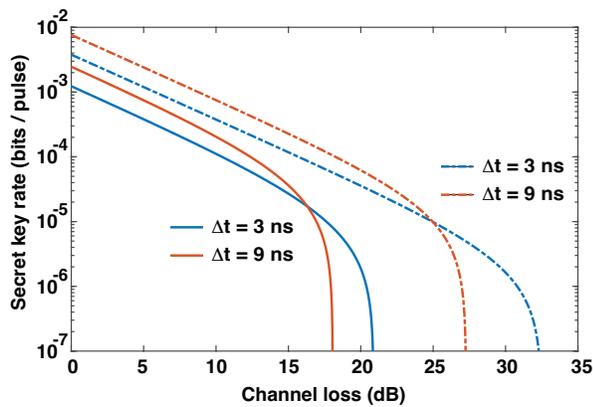}
\caption{\label{fig:fig5} Secret key rate  vs channel loss simulations for $\Delta t~=~3~ns$ and $\Delta t~=~9~ns$ values. Solid lines are obtained with the experimentally measured parameters while the dashed lines are calculated for an optimized experimental setup with better optical transmission and lower dark counts of APDs.}
\end{figure}

To give an idea about the capacity of our SPS, we simulate the secret key rate per pulse for a BB84-based QKD system with the experimentally measured parameters given above. Secret key rate is given by \citep{Waks2002a}, 
\begin{equation}
\label{rate}
R=\frac{P_{click}}{2}\{ \beta \tau(QBER)-f(QBER)h(QBER) \}.
\end{equation}
The factor $1/2$ is the efficiency of BB84 protocol and $P_{click}$ is the detection probability of a signal at Bob's side. Eq.~(\ref{rate}) also contains multi-photon probability correction $\beta=(P_{click}-P_m)/P_{click}$, where $P_m=(1/2)\mu^2g^2(0)$ is the multi-photon detection probability. Here, the expected QBER is given as 
\begin{equation}
\label{qber}
QBER=\frac{q~P_{signal}}{P_{click}}+\frac{P_{dc}/2}{P_{click}},
\end{equation}
where q is the error rate that comes from optical misalignment, $P_{signal}$ is the probability of detecting a signal, and $P_{dc}$ is the total probability of detecting dark count in a time window, which can be written as the product of the dark count rate and the temporal duration ($P_{dc}=r_d~\Delta t$). Finally, the function $\tau(e)$ is used as compression factor that accounts the photon splitting attacks, f(QBER) represents the effect of the error correction process and h(QBER) is the well-known binary Shanon-entropy. As visible in the above equations, the mean photon number of the source and the purity of the emission has a strong influence on the upper value of SKR while the dark count of detectors and the loss in the system limit the maximum distance for the useful SKR. In addition, applying a controlled amount of temporal filtering can improve the purity of the source and the QBER, therefore, extends the secret key rate for longer distances. To quantify these effects, simulations are performed for two temporal filtering window values, results of which are shown in Fig.~\ref{fig:fig5}. Here, the solid lines represent the simulations for the measured parameters of the source and the existing experimental setup: $\mu~=~0.0117$ (0.0234) and $g^2(0)~=~0.046$ (0.08) for $\Delta t~=3~ns$~(9~ns). Please note that the given mean photon number is measured right before the quantum channel, which is quite comparable with the other state-of-the-art SPSs~\citep{Kupko2020,Takemoto2015}. In addition, ~ 35\% transmission of QKD setup (including optical losses, coupling into multi-mode fibers, and coupling to the APDs), 70\% quantum efficiency of the APDs with a dark count of 1.5 kHz are considered for these simulations. Finally, to evaluate the performance of our SPS, simulations are repeated for a realistically optimized setup. A direct coupling of single photon emission to the QKD system without a PM fiber in Alice leads almost a factor of 2 enhancement in mean photon number. In addition, removing the multimode fibers and replacing the APDs with the ones that has lower dark counts in Bob side enhance the optical transmission and SKR as well as improve the QBER. Dashed lines in Fig.~\ref{fig:fig5} show the results for $\mu~=~0.023$ (0.046) and same $g^2(0)$ values as before under $\Delta t~=3~ns$~(9~ns) conditions. As observed, our source performs very well for shorter distances while it can still operate at longer distances with cost-friendly and practical single photon detectors.

In conclusion, we have presented, to the best of our knowledge, the first demonstration of QKD with single photons generated from defects in hBN at room temperature. Due to the limited bandwidth of high-voltage amplifier driving the electro-optical modulator, the QKD experiment is operated at 1 MHz, which results in a secure key rate of 238~bps and a QBER of 8.95\%. Final key rate can easily be increased up to tens of kHz by using a high speed voltage amplifier to drive the EOM or using a resonant modulator that can operate at a much lower driving voltage and consequently with high speed. Note that, the results presented in this work are obtained from a regular defect in hBN without any special technique and/or components, such as highest available NA microscope objective, to enhance the collection efficiency. Brightness of single emitters in hBN can be improved via coupling to plasmonic antennas or cavity-based structures~\citep{Tran2017b,Hausler2021}. A recent study shows that utilization of commercially available solid immersion lenses (SIL) on hBN enhances the excitation and photon collection efficiencies without the need of precise positioning of the SIL~\citep{Zeng2022}. In addition, single photon purity of the source can be improved strongly by coupling the emitters to microcavities that suppress off-resonant noise in the emission spectrum~\citep{Vogl2019} or by temporal filtering of multi-photon emission processes using amplitude modulation technique~\citep{Ates2013}. As a final note, hBN is a promising candidate for practical implementation of high-speed and long distance QKD applications because it hosts several types of defects with a bright and sharp ZPL emission over a wide spectral range from UV to NIR~\citep{Tran2017,Camphausen2020} that are attractive for long distance free-space transmission. 



This work was supported by the Scientific and Technological Research Council of Turkey (TUBITAK) grant no. 117F495 and 118F119. Authors acknowledges TUBITAK-BILGEM for providing equipment. SA acknowledges the support from Turkish Academy of Sciences (TUBA-GEBIP) and BAGEP Award of the Science Academy. 
\bibliographystyle{apsrev4-1}
\bibliography{QKD_hBN_ArXiv1}

\end{document}